\documentclass{article}
\usepackage[affil-it]{authblk}
\usepackage{graphicx}
\usepackage[space]{grffile}
\usepackage{latexsym}
\usepackage{amsfonts,amsmath,amssymb}
\usepackage{url}
\usepackage[utf8]{inputenc}
\usepackage{hyperref}
\hypersetup{colorlinks=false,pdfborder={0 0 0}}
\usepackage{textcomp}
\usepackage{longtable}
\usepackage{multirow,booktabs}

\begin{document}

\title{Quantum effects in linguistic endeavors}

\author{F.Tito Arecchi}
\affil{Professor Emeritus of Physics -University of Firenze and 
                INO-CNR-Largo E. Fermi,6, 50125,Firenze, Italy
}

\date{\today}

\bibliographystyle{unsrt}

\maketitle

\begin{abstract}
Classifying the information content of neural spike trains in a linguistic endeavor,  an uncertainty relation emerges between the bit size of a word and its duration. This uncertainty is associated with the task of 
synchronizing the spike trains of different duration representing different words. The uncertainty involves peculiar quantum features, so that word comparison amounts to measurement-based-quantum computation. Such a 
quantum behavior explains the onset and decay of the memory window connecting successive pieces of a linguistic text. The behavior here discussed is applicable to other reported evidences of quantum effects in human 
linguistic processes, so far lacking a plausible framework, since either no efforts to assign an appropriate quantum constant had been associated or speculating on microscopic processes dependent on Planck's constant 
resulted in unrealistic decoherence times.
\end{abstract}

\section{Introduction}

In cognitive science, we distinguish two moments of cognition\cite{arecchi2011phenomenology}, namely,

\textit{(A)- perception}, whereby a coherent organization emerges from the recruitment of neuronal groups, by the joint action of bottom-up sensorial stimuli and top-down perturbations arising from the long term memory\cite{grossberg1995attentive}; the joint action is adequately interpreted as a Bayes inference\cite{kersten2004object}\cite{K_rding_2006}\cite{Griffiths_2006}; perception needs an amount of time less than 1 sec\cite{rodriguez1999perception}; being a classical operation, it can be implemented by a standard computer , yielding the so-called expert systems;

and
  
\textit{  (B)-judgment}, that entails the comparison of two perceptions acquired at different times, and put into interaction via the short term memory. The judgment is the decision emerging from this comparison of two 
successive perceptions. This occurs already in experiments on animal orientation as mice in a maze\cite{Lisman_2013}. In experiments on human subjects, the sequential flashed  presentations of bistable figures entails  a 
temporary influence of the previous one upon the next; the duration of such influence is measured via a suitable K-test that yields a relevant temporal window of 2-3 sec\cite{arecchi2015test}. In the human case, language 
coding introduces an innovation in the comparison of successive pieces coded in the same language (literary, or musical or figurative). As a fact, exploring human subjects with sequences of simple word, we find evidence of a 
limited time window around 3 sec\cite{P_ppel_1997}\cite{poppel2004lost}, corresponding to the memory retrieval of a linguistic item in order to match it with the next one in a text flow. Thus, upgrading from an elementary 
bistable code\cite{arecchi2015test} to sophisticated linguistic sequences, the available time window for (B) remains about the same.
 Since (B) lasts around 3 seconds, the semantic value of the pieces under comparison must be decided within this time. This implies a fast search of the memory contents scanning all possible 
 meanings\cite{humboldt1988language} of the words in the two pieces under comparison, in order to match each-other. Speed up in a search task requires replacement of a classical algorithm with a quantum algorithm\cite{Grover_2001}\cite{Yoo_2014}.

Thus, \textit{judgment} of a linguistic sequence is the interpretation of a piece based upon the meanings of the next piece. Scanning all possible meanings of each piece entails a fast search process that requires a quantum 
search. In such a case,  a Bayes inference would not be appropriate, since the manifold of possible meanings of the previous piece has to be investigated, rather than relying on a built-in likelihood as in perception. This 
investigation has to be completed within 3 sec; otherwise, the sequence must be repeated; whence the advantage of a quantum search to speed up the search process. At the end of the comparison task, a decision center (called GWS=global workspace) picks up the information of the largest synchronized group and {\textendash} based upon it- elicits a decision\cite{baars1993cognitive}\cite{12829797}.

In this paper, we show how the comparison of two words, suitably coded as neural events, yields novel quantum features lasting long enough to affect the decision process.

\section{The physics of a linguistic task: relevant assumptions}

i)Temporal trains of identical spikes of unit area positioned at unequal times represents a sound model for the electrical activity of a cortical neuron\cite{rieke1999spikes}. The spikes have a duration 1 msec, with a minimal separation of $\tau=3\ msec$ (bin). The corresponding neuronal signal is a binary sequence of 0's and 1's, depending on whether a given bin is empty or filled by a spike. Each cortical neuron has two ways to modify its spike occurrence, namely, either coupling to other cortical neurons or receiving signals from extra-cortical regions. Thus, a meaningful linguistic piece is coded by a train of neural spikes

ii) Spike synchronization, i.e. temporal coincidence of 0{'}s and 1{'}s, is considered as the way cortical neurons organize in a collective state\cite{rieke1999spikes}\cite{Singer_1995}\cite{Arecchi_2004}\cite{Womelsdorf_2007}. In the perceptual case (A), a relevant conjecture, called {\textquotedblleft}feature binding{\textquotedblright}\cite{Singer_1995}\cite{Womelsdorf_2007}, provides a sound guess on how the spike trains in distant neuronal areas get synchronized.

iii) Any linguistic item consists of successive pieces (groups of words in literary language; groups of measures in music; small areas of a painting scanned by sequential fixations) to be compared. Precisely, a short term memory mechanism recalls the previous piece and compares it with the next one. This comparison consists of a synchronization process between the trains. If a word has \textit{N} different meanings in our private memory, the most appropriate meaning is the one that has the largest synchronization with the next piece

\begin{figure}
\begin{center}
\includegraphics[width=0.7\columnwidth]{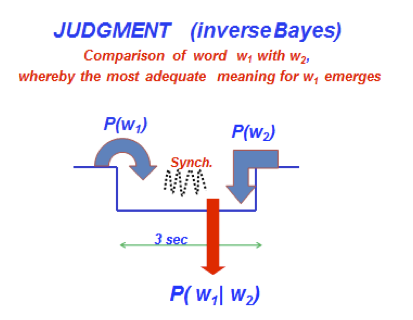}
\caption{Word meaning assignment resulting from synchronization. Ref.\cite{arecchi2011phenomenology} introduced the inverse Bayes inference by dealing with probabilities P, whereas this paper considers quantum states with probability amplitudes}
\end{center}
\end{figure}

iv) In virtue of the theta-gamma modulation of the EEG, the spike train coding the first word is interrupted from a duration $T$ to a duration $\Delta T<T$. To perform the synchronization, it must be lengthened by $T-\Delta T$ and this can occur in $N=2^{(T-\Delta T)}$ ways by filling the $T-\Delta T$ interval with $N$ different sequences of 0 and 1. Thus the first word is coded as a cluster state $|E>$ made of $N$ different sequences. We make here a bold conjecture, that the state $|E>$ is entangled, namely, that the $N$ component states have quantum-like correlations.

If e.g.    $\Delta T =T-1$, then the synchronization task of $|T>$ with $|\Delta T>$ amounts to comparing $|T>$ with the entangled state     $|E> = 1/\sqrt(2)(|\Delta T,0>+|\Delta T,1>)$,
 without further treatment, the most synchronized state will be identical to the second world. If however the $N$ states are each weighted differently by a semantic operator that we call $e$ ($e$ stays for emotion), then the most synchronized word emerging is the joint result of $e$ and of the code of the second world

\section{Uncertainty relations associated with synchronization processes}

Let us define a processing time $T$ as the time it takes for the brain to build up a complete decision like naming a picture or reading a word aloud. It corresponds  to the reaction time for visual lexical decisions or word naming, it occurs , in a range from 300 to 700 ms\cite{rodriguez1999perception}.Then, the total number of binary words that can be processed is $P_M=2^{T/\tau}$. If e.g. T=300 msec, it follows $P_M {\approx} 10^{33}$

\begin{figure}
\begin{center}
\includegraphics[width=0.7\columnwidth]{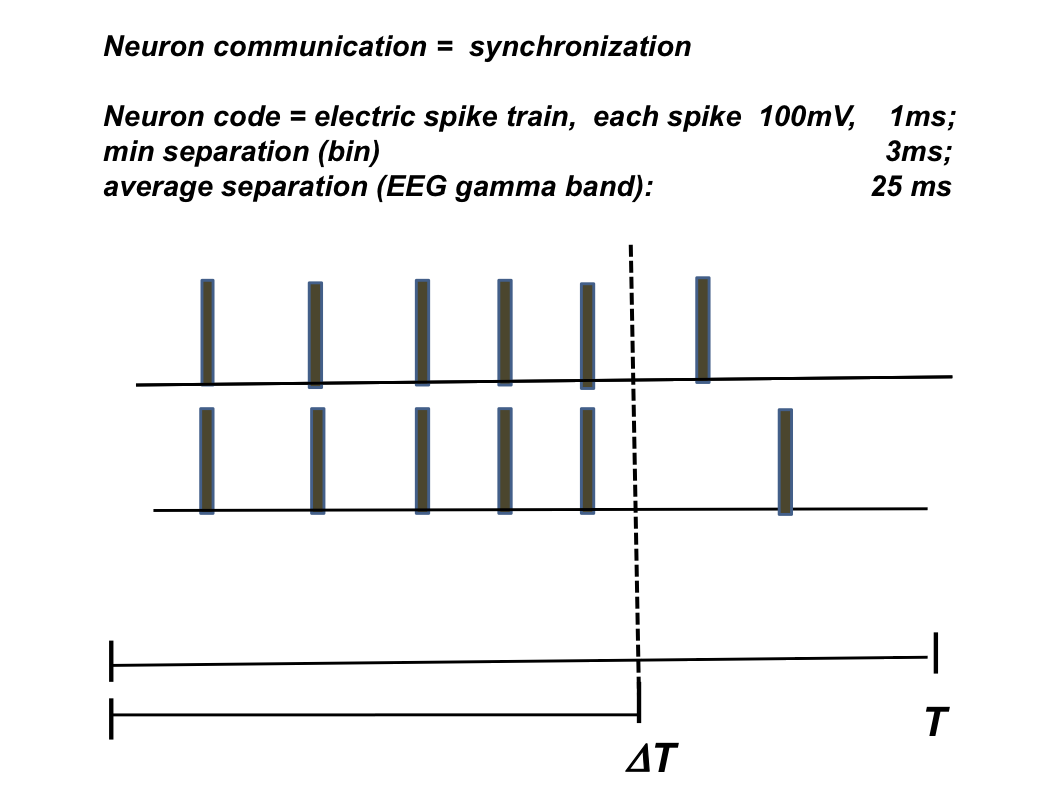}
\caption{Two spike trains of duration $T$, synchronized up to $\Delta T$. The number of different realizations is $2^{(T-\Delta T)}$%
}
\end{center}
\end{figure}

Considering the synchronization task between spike trains, we show that interruption of a spike train introduces a joint uncertainty in the word information and spike duration. Let us investigate what brain mechanisms rule the duration time. The threshold for spike onset is modulated by a EEG gamma frequency oscillation around 50 Hz; spike threshold being lowest close to the maxima of the gamma wave. Phase coherence of the gamma wave over distant brain regions permits spike synchronization overcoming delays due to the finite propagation speed in the neural axons\cite{Fries_2005} \cite{Fries_2007}. Furthermore, a lower frequency EEG oscillation (theta band , around 7 Hz) controls the number of gamma maxima within a processing interval. The theta-gamma cross-modulation corresponds to stopping the neural sequence at $\Delta T \leq T$\cite{Jensen_2007}\cite{Lisman_2013}. As a result, all spike trains equal up to $\Delta$T, but different by at least one spike in the interval $T-\Delta T$, provide an uncertainty cloud $\Delta P$ such that \cite{arecchi2003chaotic}\cite{arecchi2004uncertainty}

\begin{equation}
\Delta P= 2^{(T-\Delta T)/\tau}=P_M 2^{-\Delta T/\tau}
\end{equation}

Thus we have a peculiar uncertainty of exponential type between spike information P and duration T, that is,
\begin{equation}
\Delta P \cdot 2^{\Delta T/\tau}=P_M 
\label{eq:spike}
\end{equation}

The whole $T$ train is a vector of a $2^T$ dimensional Hilbert space; synchronization of two different $T$ trains amounts to counting the number of 0 and 1 coincidences. 

Since synchronization entails equal lengths of the trains under comparison, a $\Delta T$ pulse acquires a length $T$ by being entangled with all possible sequences of 0{'}s and 1{'}s that can be contained in the complementary interval $T-\Delta T$. 
For sake of providing an example, let us take $T=10$, $\Delta T =9$; the following synchronization values result:  $S_1=(9+1)/10=1$; $S_2=(9+0)/10=0.9$

  Define a fractional bit number $u=P/P_M$; then the fractional uncertainty $\Delta u=\Delta P/P_M$ is related to the gated time $\Delta T$ by
  
  \begin{equation}
  \Delta u \cdot 2^{\Delta T/\tau}=1
\end{equation}               

The two conjugated quantities (i.e., fractional bit number $u$ and duration of the gated spike train) are coupled by an exponential type uncertainty relation. By a change of variable

 \begin{equation}
  y= \tau 2^{t/\tau}
\end{equation}   

 where now $y$ has time dimensions , we arrive to a product type uncertainty relation
 \begin{equation}
  \Delta u \Delta y=\tau
\end{equation}   
Thus, in the space (u,y) we have a Heisenberg-like uncertainty relation. 

Historically, standard quantum physics emerged from the Newtonian dynamics of a single particle. Referring for simplicity to 1-dimension, the uncertainties of position $x$ and momentum $p$ obey the Heisenberg relation

\begin{equation}
\Delta x \Delta p\geq \hbar/2
\label{eq:indeterminazione}
\end{equation}  
All quantum facts can be derived as a consequence of Eq.(\ref{eq:indeterminazione}). Indeed, comparison with the Fourier condition   $\Delta x \Delta k\geq 1/2$  suggests the De Broglie relation 

\begin{equation}
k=p/\hbar
\end{equation}
whence the single particle interference, which contains the only quantum mystery\cite{feynman2013feynman}.

Going back to the quantization of interrupted spike trains, the associated quantum constant $C$ can be
 assigned in joules x sec, once we explicit the energy per spike. This energy corresponds to the opening along the axon of about $10^7$ ionic channels\cite{Debanne_2011} each one entailing an ATP $\rightarrow$ADP+P chemical reaction yielding 0.3 eV, thus the minimal energy disturbance in neural spike dynamics is $3 \cdot 10^{-13}J$, that is, around $10^8 k_B T_r$ , (where $k_B$ is the Boltzmann constant and $T_r$ is the room temperature ). Since $\tau=3\ msec$ , it results  $C\approx10^{-15}Js\approx 10^{19}\hbar$

However, due to the structure of Eq.(\ref{eq:spike}), the uncertainty holds over a finite range, between two extremes, namely (measuring times in $\tau$ units)

i)	$\Delta T_{min} =1$, yielding $\Delta P_{max}=P_M/2$

and 

ii)	$\Delta T_{max} =T$, yielding $\Delta P_{min}=1$

Following the standard procedure of a quantum approach, we expect single particle interference and two particle entanglement in such a space.

Once a quantum state has been prepared, it lasts over a decoherence time.

For $\Delta P=1$ (minimal disturbance represented by 1 spike) we have the decoherence time    

\begin{equation}
\Delta y_d=P_M \tau
\end{equation}

Using the numbers already reported above, the decoherence time scales as $\log_2 P_M=100 \cdot (bins)$, and going from bins to sec:
\[
\tau_d = decoherence\ time=0.3 sec               
\]

very far from the value $\tau_d=\hbar/k_B T_r \approx 10^{-14}$   evaluated for single Newtonian particles disturbed by the thermal energy $k_B T_r$\cite{Tegmark_2000}\cite{Koch_2006}.

Notice that the resulting decoherence time is equal to the full processing time $T=300 msec$ chosen as an example. If we consider a different full processing time, the decoherence time changes accordingly.

The equivalent of the De Broglie wave results as follows. Comparing Eq.(\ref{eq:indeterminazione}) with the Fourier relation $\Delta \omega \Delta t \geq 1$, we introduce a wave-like assumption
\begin{equation}
\omega=\frac{1}{\tau}\frac{P}{P_M}=\frac{u}{\tau}
\label{eq:fourierperiodicity}
\end{equation}
If $N$ is the total spike number over time $T$ and $N^{\prime}$ the spike number in the interrupted interval $\Delta T$ then
\begin{equation}
u=\frac{P}{P_M}=\frac{2^{N^{\prime}}}{2^N}=2^{-(N-N^{\prime}}=2^{-(T-\Delta T)/\tau}
\end{equation}

\section{Self interference in synchronization processes}

The equivalent of the two slit self interference of a single particle would be the comparison of a single spike train of $P$ bits with a delayed version of itself. 
A laboratory implementation would be to have the spike train translated in time and superposed to the original train. As one changes the  time separation $\Delta t$   of the two trains, the spike synchronization (number of coincidences of 1{'}s) decays as soon as   $\Delta t > \tau$   from $N$ (number of 1's in the spike train) to $\sqrt(N)$ (random overlaps).  However, further increasing the time separation, self interference entails a revival of the synchronization depending on the Fourier periodicity (Eq.\ref{eq:fourierperiodicity}), that is, for                       
\begin{equation}
\Delta t=\tau \frac{P_M}{P}=\tau 2^{(T-\Delta T)/\tau}=\tau 2^x
\end{equation}
where we call $x=(T-\Delta T)/{\tau}$  the normalized time lapse between the whole train and the interrupted version. Thus -in order to have revivals- the time translation $y=\Delta t/\tau$  must be larger than the time lapse $x$.

Comparing two interrupted sequences with lapses respectively x and x{'}, we generate two interferential returns corresponding respectively to x and x{'} and thus separated by (x{'}-x).

The wave character , that in conventional quantum mechanics is associated with $k=p/h$, here is due to the duration of the sequence to be synchronized with the initial reference of duration $T$. Thus it
seems bound to the theta {\textendash}gamma cross modulation.

In Fig.3 we present the proposal for the time equivalent of the standard two slit interference. Let us consider sequential presentations of the flashed Necker cube(fig.3a) [30]. Each presentation lasts 10ms and it is repeated each 1 sec. No switch from \#1 to \#2 (that correspond to {\textquotedblleft}1$\equiv$ face up{\textquotedblright} and {\textquotedblleft}2$\equiv$ face down{\textquotedblright}, respectively), is reported. Repeating the presentation for many observer subjects , it has been established that the threshold time for face discrimination is about $\Delta$$\tau$u-d $\cong$ 200 ms ,

Thus, if we flash the Necker cube for a time $\delta t \ll 200$ ms and then repeat the flash after a time $\Delta t \gg 200$ ms, then we should observe specific time positions where it appears ALWAYS the face UP (or DOWN), whereas elsewhere we register alternations UP-DOWN with 50\% average alternation, as illustrated in fig.3c).

\begin{figure}
\begin{center}
\includegraphics[width=0.7\columnwidth]{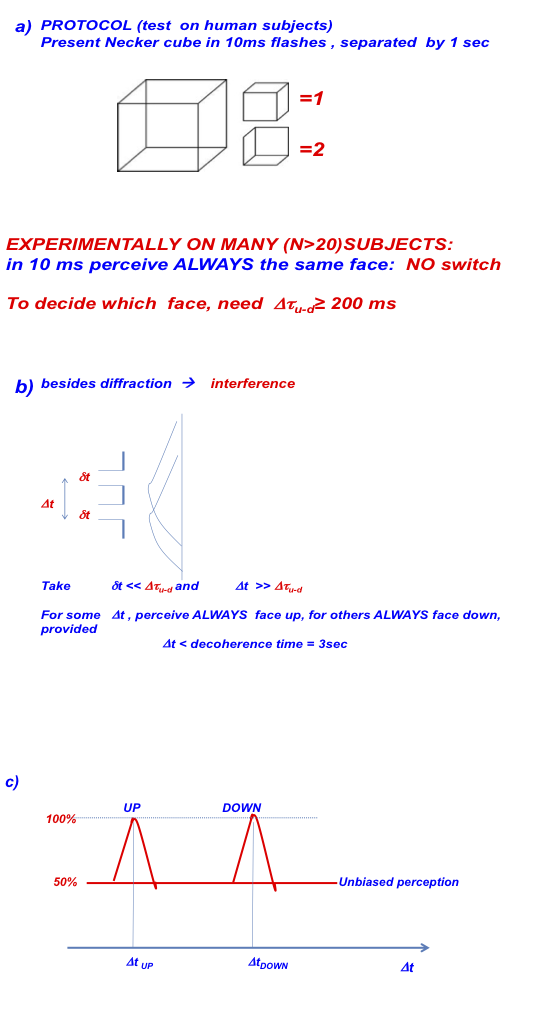}
\caption{Quantum interference in the flashed presentation of a bistable figure: a) the Necker cube;
b) flashed presentations of cube (over time $\delta t$) separated by time $\Delta T$;
c) consistent visualization of UP (DOWN) face at times $\Delta T_{UP}$ ($\Delta T_{DOWN}$)}
\end{center}
\end{figure}

\section{Quantized synchronization in a linguistic task}

Let us now apply the above formalism to a linguistic task

As said above, a linguistic task consists of the comparison of two words, one corresponding to the last presentation, and the previous one recovered by the short memory within 2-3 sec.

The words are coded as trains of neuronal spikes. Performance of the linguistic task amounts to synchronization of the two word trains. Take $T$ as the time duration of the second word. The previous one is interrupted at $\Delta T<T$ by the theta-gamma cross modulation. From what said above such a word spans a region of a functional space, that can be  taken as a finite-dimensional Hilbert space.

 The total spike train belongs to a Hilbert space of $2^T$ dimensions. It is represented by the ket $|T>$. The spike train interrupted after a duration $\Delta T$ provides a set of states living in the same $2T$-dimensional Hilbert space if it is entangled with all possible realizations of 1s and 0s in the complementary interval $T-\Delta T$. For sake of reasoning, let us consider the minimal interruption, that is $\Delta T=T-1$
 
In fact, the synchronization task of $|T>$ with $|{\Delta}T>$ amounts to comparing $|T>$ with the entangled state 
\begin{equation}
|E> = 1/\sqrt{2}(|{\Delta}T,0>+|{\Delta}T,1>)
\end{equation}
and then performing a measurement based quantum computation\cite{PhysRevLett.86.5188}\cite{briegel2009measurement}.

In general, $T-\Delta T=N$, hence synchronization amounts to measuring over a cluster of  $2^N$ entangled states, that is, comparing the whole train of duration $T$ with  $2^N$ different interrupted versions of it
each one displaying differences from the original $T$ train and hence defeating full synchronization.

Thus, we model a word recognition process as the comparison between a reference word living in a finite dimension Hilbert space of $2^T$ dimensions, and a tentative word retrieved via the short term memory and interrupted to $\Delta T$ by the theta-gamma EEG cross modulation.

If we compare with the general approach of measurement-based quantum computation , there, once a cluster (h) has been prepared, its component states must be coupled by same interaction. As a fact, the theta-gamma modulation creates the cluster (h) of entangled states; but the successive synchronization S selects state $|d>$ and thus it would be an irrelevant operation. We postulate that the {\textendash}before S is applied- the entangled states are coupled by emotional operators (e)\cite{damasio2000feeling}\cite{gibbs2005embodiment}

Thus our quantum language guess consists of the following sequence:

1.the interruption $\Delta T$ yields $2^{(T-\Delta T)}$  entangled states    $|h>$; 

2..each one of these states is modified by the emotional coupling ( e):   
\[
e |h>\rightarrow |h*>
\]

3. the synchronization S selects the state $| h*>$ that best synchronizes to $|d>$

Without \#2, the choice of $|h>$ due to $|d>$ would be a trivial operation.

  As discussed above, quantizing the spike train implies a time interruption. As a fact, spikes occur at average rates corresponding to the EEG gamma band (around 50 Hz). However, superposed to the gamma band, there is a low frequency background (theta band, around 7 Hz), that controls the number of gamma band bursts\cite{Yoo_2014}. For instance, gamma power in the hippocampus is modulated by the phase of theta oscillations during working memory retention, and the strength of this cross-frequency coupling predicts individual working memory performance\cite{Lisman_2013}.
  
  We here hypothesize that emotional effects raised by the first piece of a linguistic text induce a theta band interruption of the gamma band bursts, thus introducing a quantum feature that speeds up the exploration of the semantic space in search of the meanings that best mutually match. In this behavior, emotions do not have an esthetic value ``per se'', as instead maintained by neuro-esthetic approaches\cite{zeki1999inner}, but rather they introduce the quantum feature necessary to provide a fast scanning of all possible meanings within a decoherence time, that results to be around 3 sec. Hence the final decision does not depend on the emotions raised by the single word but it is the result of the comparison of two successive pieces of a linguistic sequence.

\begin{figure}
\begin{center}
\includegraphics[width=0.7\columnwidth]{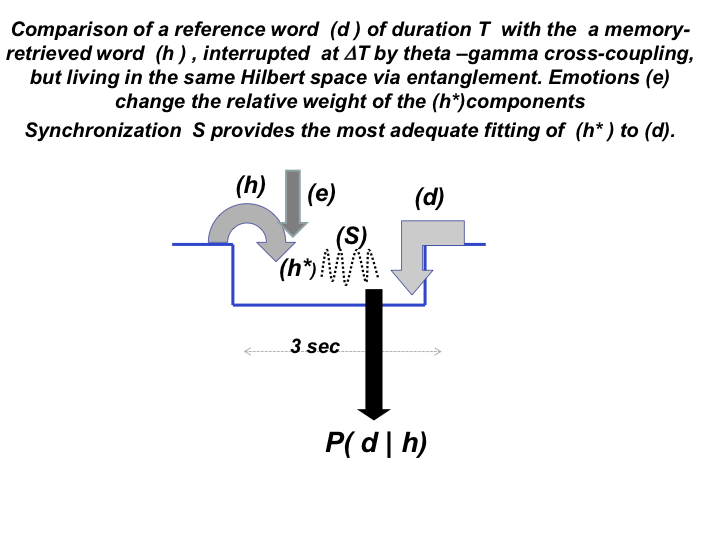}
\caption{The procedure conjectured in linguistic endeavors- A previous piece of a text ($h$) is retrieved by the short term memory, modified by emotions ($e$ )into ($h^{\star}$) and compared via synchronization ($S$) with the next piece ($d$). The most adequate fit $P(d|h)$ emerges as a result of the comparison (judgment and consequent decision)[inverse Bayes procedure].
}
\end{center}
\end{figure}

\section{Main features of the physics of a linguistic endeavour}

    To conclude, we stress the revolution brought about by the linguistic processes in human cognitive endeavors, namely, a new type of quantum behavior has to be considered; the interrupted spike synchronization is a peculiar physical process that cannot be grasped in terms of Newtonian position-momentum variables; hence, the quantum constant for spike train position-duration uncertainty has nothing to do with Planck{'}s constant;
 
 The minimal energy disturbance which rules the decoherence time is by no means $k_B T_R$ ($T_R$ being the room temperature); rather, since it corresponds to $\Delta P=1$, it entails the minimal energy necessary to add or destroy a cortical spike. This energy corresponds to the opening along the axon of about $10^7$ ionic channels each one requiring an ATP $\rightarrow$ADP+P chemical reaction involving 0.3 eV, thus the minimal energy disturbance in neural spike dynamics is around $10^8$ $k_B T_R$. This is the evolutionary advantage of a brain, that is, to live comfortably at TR and yet be barely disturbed, as it were cooled at $10^{-8}$ the room temperature.
 
 The procedure here described for connecting two successive pieces of a linguistic text is applicable to other reported evidences of quantum effects in human cognitive processes, so far lacking a plausible framework since no efforts to assign a quantum constant have been associated. 
Models of quantum behavior in language and decision taking have already been considered by several Authors but without a dynamical basis, starting 1995\cite{Aerts_1995}\cite{Aerts_2009} and over the past decade\cite{khrennikov2010ubiquitous}. Most references are collected in a recent book\cite{busemeyer2012quantum}.
The speculations introduced to justify a quantum speed up can be grouped in two categories , depending on how they do not fulfill the requirements of a rigorous quantum approach, that is,
i)	either they lack a dynamical basis and thus do not discuss limitations due to a quantum constant, hence, they do not inquire for a decoherence time terminating the quantum operation or
ii) they refer to the quantum behavior of Newtonian particles\cite{penrose1994shadows}\cite{hagan2002quantum}and hence are limited by a decoherence time estimated around $10^{-14}$ sec,  well below the infra-sec scale of the cognitive processes.

    To summarize the above considerations, a quantum behavior entails pairs of incompatible observables, whose measurement uncertainties are bound by a quantum constant. One cannot apply a quantum formalism without having specified the quantum constant ruling the formalism. Our approach provides evidence of a new quantum behavior associated with the synchronization of time limited sequences of spikes; the emerging decoherence time is compatible with the observed processing times in linguistic endeavors. On the contrary, all previously reported approaches either overlook the need for a quantization constant , or they quantize Newtonian particles by using Planck constant and consequently arrive to very short decoherence times, incompatible with human linguistic processes.
We have seen that, while in the perceptual case, the cognitive action combines a bottom-up signal provided by the sensorial organs with a top-down interpretation provided by long term memories stored in extra-cortical areas, in the linguistic case, the comparison occurs between the code of the second piece and the code of the previous one retrieved by the short term memory. In this second case , theta {\textendash}gamma cross modulation  introduces a quantum uncertainty, hence an entanglement among different words that provides a fast quantum search of meanings.

\section{Acknowledgments}
I am extremely grateful to Augusto Smerzi for critical remarks and helpful suggestions. I am indebted to Alessandro Farini, with whom I am exploring different experimental implementations of self-interference

\bibliography{bibliografiatito.bib}

\end{document}